\documentclass[aps, 
prl,
twocolumn,
superscriptaddress,
showpacs,preprintnumbers,amsmath,amssymb,
longbibliography,floatfix,10pt]{revtex4-2}

\usepackage[
colorlinks=true, 
pdfstartview=FitV, 
linkcolor=blue, 
citecolor=magenta, 
urlcolor=blue, 
bookmarks=true,
bookmarksnumbered=true,
pdftitle={},
pdfauthor={}
]{hyperref}

\usepackage[dvips]{graphicx}
\usepackage{bm,latexsym,amsmath,amssymb,amsfonts,mathrsfs,textcomp}
\usepackage{color}
\input{colordvi.tex}
\usepackage{comment}
\usepackage{simpler-wick}
\usepackage{tikz}
\usepackage[compat=1.1.0]{tikz-feynhand}
\usepackage[normalem]{ulem}

\newcommand\sect[1]{{\it #1.}---}

\newcommand{\rmi}{\mathrm{i}} 
\newcommand{\rme}{\mathrm{e}} 
\newcommand{\rmd}{\mathrm{d}}

\newcommand{\for}{\textrm{for}}

\newcommand{\eff}{\textrm{eff}}

\newcommand{\inj}{\textrm{in}}

\newcommand{\tra}{\textrm{tr}}
\newcommand{\refl}{\textrm{re}}

\newcommand{\bx}{\bm{x}}

\newcommand{\bnab}{\bm{\nabla}}

\newcommand{\calH}{\mathcal{H}}

\newcommand{\calQ}{\mathcal{Q}}

\newcommand{\dis}{\displaystyle}

\newcommand{\average}[1]{\langle#1\rangle}

\newcommand{\cout}[1]{ \if 0 {#1} \fi }

\begin{document}

\title{
Anomalous tunneling as a low-energy theorem for Nambu-Goldstone modes
}

\author{Keisuke Fujii}
\email{fujii@phys.sci.isct.ac.jp}
\affiliation{Department of Physics, Institute of Science Tokyo,
Ookayama, Meguro-ku, Tokyo 152-8551, Japan}

\author{Daichi Kagamihara}
\email{dkagamihara119@g.chuo-u.ac.jp}
\affiliation{Department of Physics, Chuo University, Bunkyo, Tokyo 112-8551, Japan}

\author{Masaru Hongo}
\email{hongo@phys.sc.niigata-u.ac.jp}
\affiliation{Department of Physics, Niigata University, Niigata 950-2181, Japan}
\affiliation{RIKEN Center for Interdisciplinary Theoretical and Mathematical Sciences (iTHEMS), RIKEN, Wako 351-0198, Japan}

\begin{abstract}
Anomalous tunneling refers to the phenomenon in which the transmission coefficient through a potential barrier approaches unity as the energy of an incident particle or quasiparticle tends to zero. 
This counterintuitive effect has been reported in systems exhibiting spontaneous symmetry breaking (SSB), such as superfluids, yet the general conditions for its occurrence remain unclear. 
In this Letter, we establish that anomalous tunneling of Nambu-Goldstone (NG) modes is a universal low-energy theorem dictated solely by symmetry and scaling, using a low-energy effective field theory (EFT) framework.
We formulate the scattering of NG modes by external potentials in terms of spatially dependent EFT coefficients and demonstrate that symmetry-preserving localized potentials are irrelevant in the long-wavelength limit, leading to perfect transmission. 
In contrast, symmetry-breaking perturbations are relevant and suppress transmission, resulting in the absence of anomalous tunneling.
We illustrate this universal behavior with explicit examples of superfluid phonons and magnons.

\end{abstract}
\maketitle

\sect{Introduction}%
\label{sec:intro}%
Quantum tunneling is a quintessential phenomenon absent in classical physics.
It allows quasiparticles with energies below a potential barrier to pass through the barrier.
Tunneling plays a crucial role in a wide range of phenomena, from fundamental processes in physics such as alpha decay~\cite{Gamow:1928,gurney1928wave} and macroscopic quantum tunneling in superconducting systems to chemical reactions~\cite{ChemicalReactions1,ChemicalReactions2} and biological systems~\cite{Biology}.
It also underlies essential technologies, including tunnel diodes~\cite{Esaki:1958} and scanning tunneling microscopes~\cite{Binnig:1982}.

One fundamental quantity characterizing tunneling is the transmission coefficient, which typically decreases toward zero as the energy of incident particles is lowered.
In contrast, systems with spontaneous symmetry breaking (SSB), such as superfluids, exhibit the opposite behavior: the transmission coefficient approaches unity in the low-energy limit.
This counterintuitive phenomenon, known as \textit{anomalous tunneling}, was first predicted in Bose-Einstein condensates (BECs)~\cite{Kagan:2003}.

Although anomalous tunneling has been extensively studied theoretically in BECs and other systems with SSB~\cite{Kagan:2003,Danshita:2005a,Danshita:2005b,Rapedius:2006,Danshita:2006,Danshita:2007,Kato:2008,Watabe:2008,Jian-Wen:2008,Ohashi:2008,Tsuchiya:2008,Takahashi:2009,Tsuchiya:2009,Watabe:2010,Takahashi:2010,Watabe:2011a,Watabe:2011b,Watabe:2011c,Kato:2012,Watabe:2015,Nakayama:2015,Ohashi:2021}, direct experimental observation remains challenging.
BECs of weakly interacting Bose gases provide promising platforms, as both the creation of low-energy excitations~\cite{Andrews:1997aa,Ville:2018} and versatile barrier manipulation, such as Josephson junctions~\cite{Albiez:2005,LeBlanc:2011}, have been realized.
Indeed, indirect experimental evidence for anomalous tunneling has been reported in this system~\cite{kagamihara2025}.
Recent advances in matter-wave circuits~\cite{chien2015quantum,Atomtroincs:2021} enable detailed investigations of tunneling phenomena, including point contacts~\cite{Brantut:2012,Husmann:2015,Hausler:2017,Krinner_2017} and Josephson junctions in fermionic superfluids~\cite{Valtolia:2015,Kwon:2020}.
These developments pave the way for direct experimental observation and motivate a unified theoretical framework for anomalous tunneling.

\begin{figure}[t]
 \centering
\includegraphics[width=1.0\linewidth]{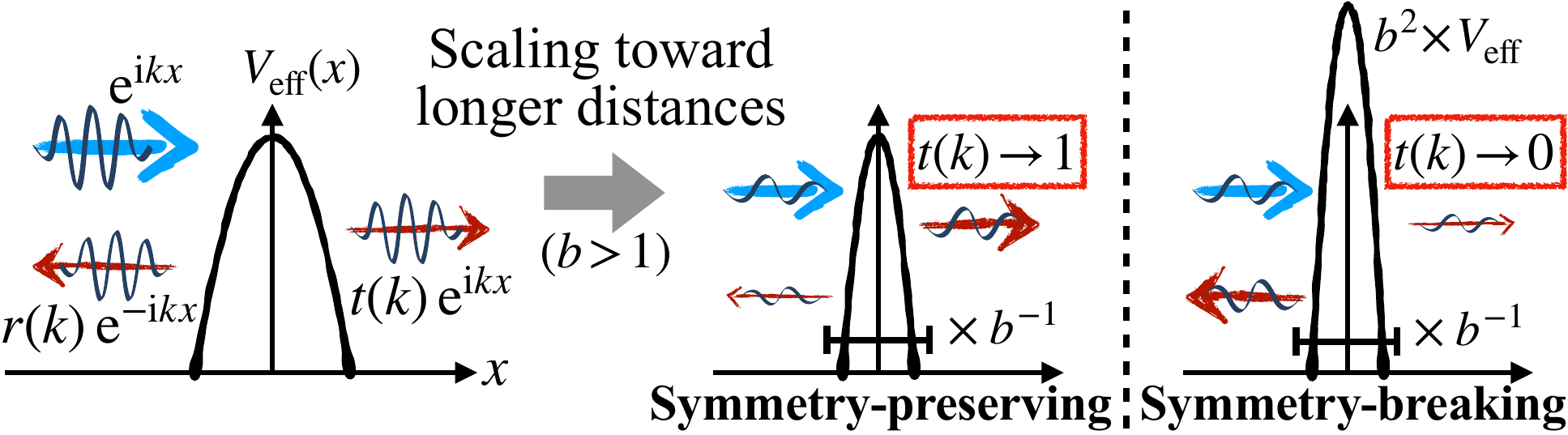}
\caption{
Schematic illustration of our result.
At low energies, the problem reduces to scattering of the NG mode in an effective potential $V_{\mathrm{eff}}(x)$.
Under scaling, symmetry-preserving potentials shrink, leading to perfect transmission, whereas symmetry-breaking ones grow, leading to perfect reflection.
}
 \label{fig:schematic}
\end{figure}

Despite extensive studies, a unified understanding of anomalous tunneling in the low-energy regime and the conditions for its occurrence is still lacking.
In this Letter, we address this issue using a low-energy effective field theory (EFT) approach.
Low-energy excitations in systems with SSB are governed by Nambu-Goldstone (NG) bosons~\cite{Goldstone:1962es,Goldstone:1961eq,Nambu:1961tp}, whose EFTs can be systematically constructed from symmetry principles without relying on microscopic details~\cite{Coleman:1969,Callan:1969,Leutwyler:1994,Watanabe:2012,Watanabe:2014}.
Within this framework, we formulate the scattering of NG modes by external potentials in terms of spatially dependent EFT coefficients.
We establish that anomalous tunneling follows as a universal low-energy theorem dictated by symmetry and scaling.
A scaling analysis then shows that symmetry-preserving potentials are irrelevant in the long-wavelength limit, leading to perfect transmission, whereas symmetry-breaking perturbations are relevant and suppress transmission (Fig.~\ref{fig:schematic}).

\sect{Setup}
Suppose that the system spontaneously breaks the internal symmetry $G$ down to its unbroken subgroup $H$.
The low-energy scattering of the associated NG modes $\pi^a(t,\bx)\,(a=1,\cdots,\dim G/H)$ from a static $G$-invariant external potential is governed by the leading-order quadratic action~\cite{Coleman:1969,Callan:1969,Leutwyler:1994,Watanabe:2012,Watanabe:2014}
\begin{align}
 S_{\eff}
 = \int \rmd t \rmd^d x 
 \bigg[
 &\frac{1}{2} g^t_{ab} (x) \partial_t \pi^a \partial_t \pi^b 
 - \frac{1}{2} g^s_{ab} (x) \bnab \pi^a \cdot \bnab \pi^b
 \notag \\
 &- \frac{1}{2} \rho_{ab} (x) \pi^a \partial_t \pi^b
 \bigg],
 \label{eq:effective-action}
\end{align}
where $g^t_{ab}(x)$ and $g^s_{ab}(x)$ denote the local susceptibility and generalized rigidity, respectively, and $\rho_{ab}(x)$ is the Watanabe-Brauner matrix characterizing type-B NG modes.
To discuss one-dimensional scattering along the $x$-direction induced by symmetry-preserving potentials under $G$, we allow all coefficient functions to depend on $x$ and approach constants at spatial infinity.

When $\rho_{ab}=0~(\forall a,b)$, the system describes a set of decoupled type-A NG modes, in which case we may focus on a representative component, e.g., $\pi^1$.
When $\rho_{ab}\neq 0~(\text{e.g., }a,b=1,2)$, the fields $\pi^1$ and $\pi^2$ form a canonical pair corresponding to a type-B NG mode.
To describe one-dimensional scattering, we consider stationary solutions with frequency $\omega$ and introduce a complex-valued mode function $\phi(x)$ as
\begin{subequations}
 \begin{align}
 &\mathrm{Type\text{-}A~case:}~
 \pi^1(t,\bx)
 = \rme^{-\rmi \omega t} \phi(x),
 \label{eq:type-A-basis}
 \\
 &\mathrm{Type\text{-}B~case:}~
 \frac{\pi^1(t,\bx) + \rmi \pi^2 (t,\bx)}{\sqrt{2}}
 = \rme^{-\rmi \omega t} \phi(x).
 \label{eq:type-B-basis}
 \end{align}
\end{subequations}
The incoming NG mode along the $+x$ direction is described by $\phi_{\inj} = \rme^{\rmi k x}$.
Including reflected and transmitted components, $\phi_{\refl} = r(k) \rme^{-\rmi k x}$ and $\phi_{\tra} = t(k) \rme^{\rmi k x}$, the solution takes the asymptotic form
\begin{equation}
 \phi(x) \to
 \begin{cases}
  \rme^{\rmi k x} + r(k) \rme^{-\rmi k x} 
  &(x \to - \infty), 
  \\
  t(k) \rme^{\rmi k x}
  &(x \to + \infty),
  \end{cases}
 \label{eq:ansatz-solution}
\end{equation}
where $k$ is determined by the asymptotic dispersion relations for $\omega$.
Then, the scattering problem reduces to determining $r(k),t(k) \in \mathbb{C}$ from the equation of motion:
\begin{subequations}\label{eq:eom}
 \begin{align}
  0 &=  \partial_x [g^s (x) \partial_x \phi (x)] 
  + \omega^2 g^t (x) \phi (x),
 \label{eq:type-A-eom}
  \\
  0 &= \partial_x [g^s (x) \partial_x \phi (x)]
 + [\omega^2 g^t (x) + \omega \rho (x)] \phi (x),
 \label{eq:type-B-eom}
 \end{align}
\end{subequations}
depending on whether the mode is type-A or type-B.
Here, we assume $g_{ab}^{s,t} = g^{s,t} \delta_{ab}$ and 
$\rho_{ab} = \rho \epsilon_{ab}$, focusing on a single irreducible sector.
We will later solve these equations in specific examples of a superfluid phonon and magnon.

To define the transmission and reflection coefficients, $T$ and $R$, we rely on the asymptotic behavior of the energy flux $\calQ^i$.
Time-translation symmetry implies that the Hamiltonian density $\calH_{\eff}$ satisfies the continuity equation $\partial_t \calH_{\eff} + \partial_i \calQ^i = 0$, where the associated Noether current is given by
\begin{equation}
 \calQ^i (t,\bx)
 = g_{ab}^s (x) \partial_t \pi^a (t,\bx) 
 \partial_i \pi^b (t,\bx).
\end{equation}
We define the transmission and reflection coefficients as
\begin{subequations}
\label{eq:TR-coefficients}
\begin{align}
 T
 &:=
 \frac{|\average{\calQ^x_{\tra}(t,x)}|}{|\average{\calQ^x_{\inj}(t,x)}|}
 = |t(k)|^2,
 \\
 R
 &:=
 \frac{|\average{\calQ^x_{\refl}(t,x)}|}{|\average{\calQ^x_{\inj}(t,x)}|}
 = |r(k)|^2,
\end{align}
\end{subequations}
where the relation $R = 1 - T$ follows from the conservation of energy flux.

\sect{Anomalous tunneling as a low-energy theorem}
We demonstrate that anomalous tunneling of NG modes is a universal low-energy theorem dictated by symmetry and scaling.
For a symmetry-preserving localized potential, the transmission coefficient approaches unity as $\omega \to 0$, while symmetry-breaking perturbations lead to $T \to 0$.
The derivation is based on a scaling analysis that generalizes the standard power counting for the coupling constant.

We start from the effective action~\eqref{eq:effective-action}, in which the effect of the external potential is entirely encoded in the coefficient functions.
For symmetry-preserving potentials, these coefficients can be decomposed as
\begin{equation}
 g^{s,t} (x)=\bar{g}^{s,t} +\delta g^{s,t} (x), \quad 
 \rho (x)= \bar{\rho} +\delta \rho (x),\label{eq:coeff-decomposition}
\end{equation}
where $\delta g^t(x)$, $\delta g^s(x)$, and $\delta \rho(x)$ are localized within a finite spatial region $\Omega(\bx)$.
The potential scattering is induced by the localized part of the action
\begin{align}
 S_{\mathrm{pot}}
 = \int_{\Omega (\bx)} \rmd t \rmd^d x 
 &\bigg[
 \frac{1}{2} \delta g^t (x) (\partial_t \pi^a)^2
 - \frac{1}{2} \delta g^s (x) (\bnab \pi^a)^2
 \notag \\
 &- \frac{1}{2} \delta \rho (x)
 (\pi^1 \partial_t \pi^2 - \pi^2 \partial_t \pi^1)
 \bigg],
 \label{eq:action-potential}
\end{align}
which has support only in $\Omega(\bx)$.

To determine the low-energy behavior, we perform a long-wavelength scaling analysis.
Outside $\Omega(\bx)$, the action reduces to that of the NG mode in a uniform system, so that
we introduce the scaling transformation
\begin{equation}
 t \to b^{-z} t, 
 \quad 
 \bx \to b^{-1} \bx 
 \quad 
 (b>1),
\end{equation}
with $z=1$ for type-A modes and $z=2$ for type-B modes.
Under this transformation, momentum scales as $k \to bk$.
Thus, keeping the rescaled momentum fixed corresponds to probing longer wavelengths in the original variables, and the limit $b\to\infty$ describes the low-momentum limit $k\to0$.
The scaling of the fields and coefficients is fixed by requiring invariance of the effective action, leading to
\begin{subequations}\label{eq:scaling-tr}
 \begin{align}
 &\mathrm{Type\text{-}A~case:}~
 \begin{cases}
  \pi(t,\bx) \to b^{(d-1)/2} \pi(b t, b\bx),
  \\
  g^{t} (x) \to g^{t} (bx),
  \\
  g^{s} (x) \to g^{s} (bx),
 \end{cases}
 \\
 &\mathrm{Type\text{-}B~case:}~
 \begin{cases}
  \pi(t,\bx) \to b^{d/2} \pi(b^2 t, b\bx),
  \\
  g^{t} (x) \to b^{-2} g^{t} (bx),
  \\
  g^{s} (x) \to g^{s} (bx),
  \\
  \rho (x) \to \rho (bx).
 \end{cases}
 \end{align}
\end{subequations}

The key observation is that the support of the localized coefficients effectively shrinks under the scaling transformation while their magnitude remains unchanged.
In particular, the one-dimensional support for a potential localized along the $x$ direction reduces its size by a factor of $1/b$.
Since the coefficients themselves do not grow under Eq.~\eqref{eq:scaling-tr}, the localized action scales as
\begin{equation}
 S_{\mathrm{pot}} \to b^{-1} S_{\mathrm{pot}}.\label{eq:scaling-Spot}
\end{equation}
Thus, the localized potential is subleading under the long-wavelength scaling.

As a consequence, a symmetry-preserving localized potential does not contribute to leading low-energy scattering.
The transmission coefficient therefore approaches unity in the low-energy limit, $T \to 1$ and $R \to 0$, thereby proving anomalous tunneling.

\sect{Universal scaling in low-energy scattering}
The scaling analysis presented above generalizes the standard power counting argument.
Within this framework, the scaling of localized couplings is governed by the scaling of the spatial support of the potential.

To make this explicit, we consider the localized limit of the potential,
\begin{equation}
 \delta g^{s,t} (x) = g_{\mathrm{loc}}^{s,t} \delta (x) , \quad 
 \delta \rho (x) = \rho_{\mathrm{loc}} \delta (x) .
\end{equation}
In this limit, the localized perturbation reduces to a point-like operator insertion, and standard dimensional power counting applies.
Assigning scaling dimension $[k]=1$, we have $[\delta(x)]=1$.
This implies $[g_{\mathrm{loc}}^{s,t}]=-1$ and $[\rho_{\mathrm{loc}}]=-1$.
Thus, these operators are irrelevant, and their contribution to scattering is suppressed in the low-momentum limit.
Moreover, this argument generalizes to defects of lower codimension: compared to plane-like defects, symmetry-preserving perturbations localized on lines or points are further suppressed at low energies.
The above dimensional analysis demonstrates that the conventional power counting follows from the scaling transformation in Eq.~\eqref{eq:scaling-tr}, as $\delta(bx)=b^{-1}\delta(x)$.

This perspective allows us to determine the universal low-momentum behavior of the reflection coefficient.
At leading order, the scattering amplitude is linear in the localized couplings.
Since $[g_{\mathrm{loc}}^{s,t}] = -1$ and $[\rho_{\mathrm{loc}}] = -1$, dimensional analysis implies that the amplitude scales as $k$.
Therefore, the reflection coefficient, being the squared amplitude, behaves as
\begin{equation}
 R \sim k^2 \quad (k \sim 0).
 \label{eq:R-coefficient}
\end{equation}
This establishes the universal low-energy scaling of the reflection coefficient.

We explicitly confirm the above scaling prediction within low-energy perturbation theory.
Defining a new coordinate $X(x):=\int^x \rmd x^{\prime} \bar{g}^s/g^s(x^\prime)$, we rewrite the equation of motion~\eqref{eq:eom} into a Schr\"odinger-type equation
\begin{align}
 [- \partial_X^2 + V_{\eff}(X) ] \phi(X) 
 = k^2 \phi(X) , 
\label{eq:eom-decomposed}
\end{align}
with the effective potential $V_{\eff}(X) := k^2 v(X)$, where
\begin{align}
 v (X) :=
 \begin{cases}
  \dis 1-\frac{g^t(x)g^s(x)}{\bar{g}^t\bar{g}^s} & (\textrm{Type-A}), 
  \vspace{3pt} \\
  \dis 1-\frac{\rho(x) g^s (x)}{\bar{\rho}\bar{g}^s} & (\textrm{Type-B}).
\end{cases}
\end{align}
Here, we used the dispersion relations $\omega^2=\bar{g}^s k^2/\bar{g}^t$ for type-A NG modes and $\omega=\bar{g}^s k^2/\bar{\rho}$ for type-B NG modes.
For the type-B case, we neglect the subleading $\omega^2g^t(x)$ term to focus on the low-energy behavior.

The crucial point is that the effective potential $V_{\eff}(X)$ is proportional to $k^2$ and thus suppressed at low momentum.
Accordingly, the standard Born approximation is justified and gives
\begin{align}
&\phi(x)\simeq \rme^{\rmi kX(x)} 
-\int \rmd X^{\prime}
G_k(X-X^\prime) k^2 v(X^\prime)
\rme^{\rmi k X^{\prime}},
\end{align}
where $G_k(X-X^{\prime})=\rme^{\rmi k|X-X^{\prime}|}/(2\rmi k)$ is the outgoing Green's function for Eq.~\eqref{eq:eom-decomposed}.

From the asymptotic form of the solution as $x \to - \infty$, we obtain the reflection amplitude as
\begin{equation}
r(k)=\frac{\rmi k}{2}\int \rmd x\frac{\bar{g}^s}{g^s(x)} v(x)+O(k^2),
\end{equation}
where we used $X(x)=x+\mathrm{const}.$ in the asymptotic region and $\rmd X=\rmd x\,\bar{g}^s/g^s(x)$ in the integral.
Therefore, we obtain the universal low-energy behavior of the reflection coefficient, including its prefactor,
\begin{equation}
R=\frac{k^2}{4}\left(\int \rmd x\,
\frac{\bar{g}^s}{g^s(x)}
v(x)\right)^2+O(k^3).
\label{eq:low-energy-behavior}
\end{equation}

\sect{Absence of anomalous tunneling}
We generalize our analysis to show the absence of anomalous tunneling in symmetry-breaking potentials.
In the simplest case, we have an additional potential term
\begin{align}
 S_{\mathrm{pot}'}
 = - \frac{1}{2} \int_{\Omega (\bx)} \rmd t \rmd^d x\,
 \Delta(x) (\pi^a)^2,
 \label{eq:action-potential-breaking}
\end{align}
which again has support only in $\Omega(\bx)$.
Under the scaling transformation \eqref{eq:scaling-tr}, this potential transforms as
\begin{equation} 
 \Delta (x) \to b^2 \Delta (bx),\label{eq:scaling-symmetry-breaking-potential}
\end{equation}
which carries a positive scaling dimension.
Under this scaling, the shrinking of the support competes with the enhancement of the potential magnitude.
In one dimension, the $b^2$ enhancement renders the perturbation relevant, thereby eliminating anomalous tunneling.

The same low-energy structure also follows from the Schr\"{o}dinger-type equation of motion~\eqref{eq:eom-decomposed}.
In this case, the effective potential is not proportional to $k^2$ at low energy, in contrast to the symmetry-preserving case.
The equation therefore reduces to an ordinary one-dimensional Schr\"{o}dinger problem with a finite short-range potential.
Accordingly, in the low-energy limit $k \to 0$, the transmission coefficient vanishes, $T \to 0$, unless the potential supports a zero-energy resonance~\cite{Newton:Scattering}.

\sect{Examples}
We discuss two representative examples: a superfluid phonon and magnons.
\paragraph{a) Superfluid phonon}
As a concrete realization of our result, we consider a nonrelativistic attractive Fermi gas or a nonrelativistic Bose gas under an external potential $V(\bx)$.
These systems exhibit superfluidity, whose low-energy behavior is universally described by~\cite{Greiter:1989,Son:2005}
\begin{align}
 S_{\eff} = \int \rmd t \rmd^d x\, p\big(\theta(t,\bx)\big),
\end{align}
with the pressure function $p$, whose argument is given by $\theta(t,\bx) := \mu-V(\bx)-\partial_{t}\pi(t,\bx)-[\bnab\pi(t,\bx)]^{2}/2m$ with the chemical potential $\mu$ and mass $m$ of the constituent particles.
This EFT relies on a local-density approximation, which requires $\mu > V(x)$.
Expanding the effective action in terms of the phonon field $\pi (t,\bx)$, we find Eq.~\eqref{eq:effective-action} with 
\begin{subequations}
 \begin{align}
  g^t (x) 
  &= \chi(\bx)
  := p^{\prime\prime} \big(\mu-V(\bx)\big), 
  \\
   g^s (x) 
  &= \frac{n(\bx)}{m}
  := \frac{1}{m} p^{\prime} \big(\mu-V(\bx)\big), 
 \end{align}
\end{subequations}
and $\rho = 0$.
Thus, the inhomogeneity induced by $V(\bx)$ is entirely captured by $g^t(x)$ and $g^s(x)$ at low energy.

Let us consider a concrete example of the scattering problem by specifying
\begin{align}
V(\bx)=\left\{\!
\begin{array}{cc}
V_{0} & \quad\for\quad |x|<a, \\
0 & \quad\for\quad |x|>a,
\end{array}
\right.\label{eq:potential}
\end{align}
In this setup, one finds the analytic solution under appropriate matching conditions at $x=\pm a$; e.g., the continuity of the phonon field $\varphi(a+0^{+})=\varphi(a-0^{+})$, and continuity of the mass flux $n_{o}\partial_{x}\varphi(a+0^{+})=n_{i}\partial_{x}\varphi(a-0^{+})$ at $x=a$.
This results in 
\begin{equation}
 T =|t(k)|^{2}=\frac{4\gamma^{2}}{4\gamma^{2}+(\gamma^{2}-1)^{2}\sin^{2}(2k_{i}a)}, 
\label{eq:T-coefficient}
\end{equation}
where we introduced quantities with a subscript $o(i)$ evaluated at the outside (inside) of the potential.
We also defined the momentums $k := \omega/c$, the sound velocities $c :=\sqrt{n/m\chi}$ for each region, and the dimensionless parameter $\gamma := \sqrt{n_{i}\chi_{i}/n_{o}\chi_{o}}$.

\begin{figure}[t]
 \centering
\includegraphics[width=0.8\linewidth]{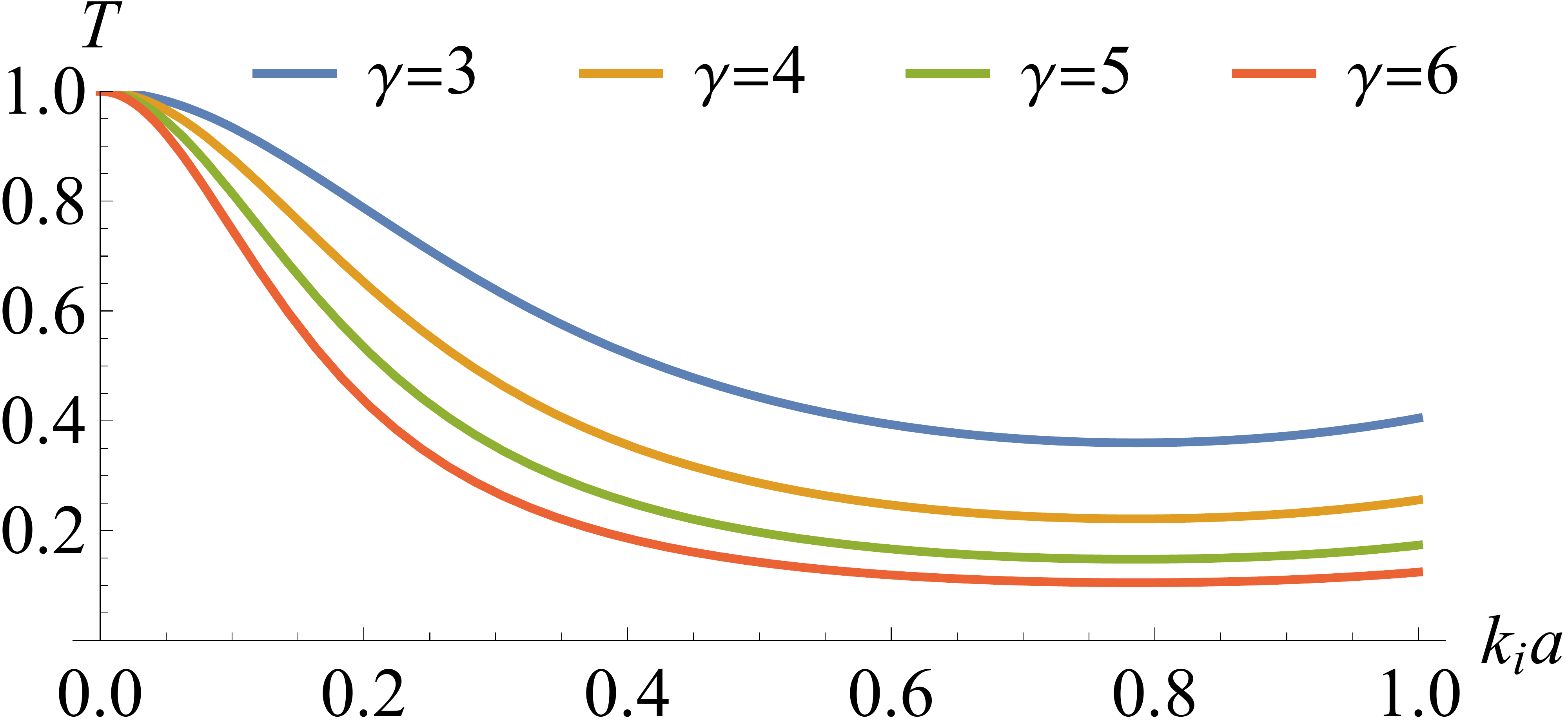}
\caption{
Transmission coefficient $T$ in Eq.~\eqref{eq:T-coefficient} as a function of $k_{i}a$ for several values of $\gamma$.
The momentum inside the barrier is related to the incident momentum by $k_{i} = (c_{o}/c_{i}) k_{o}$.
}
 \label{fig:transmission}
\end{figure}

Figure~\ref{fig:transmission} shows the transmission coefficient as a function of $k_i a$.
We find perfect transmission at $k_{i}a=0$, showing anomalous tunneling. 
Moreover, Eq.~\eqref{eq:T-coefficient} leads to
\begin{equation}
 R= 1-T
 = \biggl(\frac{n_{o}}{n_{i}}-\frac{\chi_{i}}{\chi_{o}}\biggr)^{2}(k_{o}a)^{2}
 +O\big((k_{o}a)^{4}\big),
\label{eq:R-coefficient}
\end{equation}
which confirms our formula~\eqref{eq:low-energy-behavior}.
The coefficient of the $(k_{o}a)^{2}$ term is fully determined by the equation of state, i.e., the pressure function $p(\mu)$.
For example, for a Fermi gas with an equation of state $p(\mu)\propto \mu^{5/2}$, which includes both free and unitary Fermi gases, we obtain
\begin{equation}
 \biggl(\frac{n_{o}}{n_{i}}-\frac{\chi_{i}}{\chi_{o}}\biggr)^{2}
 = \frac{v^2 (2-v)^2}{(1-v)^3} 
 \quad \mathrm{with} \quad 
 v := \frac{V_0}{\mu}.
\label{eq:coeff-example}
\end{equation}
This yields a direct experimental signature of the low-energy theorem.

\paragraph{b) Magnon in ferromagnets}
As another illustrative example, we consider magnons in a Heisenberg ferromagnet, where the ground state spontaneously breaks the spin-rotational symmetry as $\mathrm{SO}(3)\to \mathrm{SO}(2)$.
The low-energy excitations are magnons, i.e., NG modes associated with this symmetry breaking.

We introduce two types of localized perturbations within a finite region $\Omega(\bx)$:
(i) a modulation of the exchange coupling, and (ii) a magnetic field, where (i) is symmetry-preserving and (ii) is symmetry-breaking.

According to the above criterion, the transmission coefficient approaches unity for symmetry-preserving perturbations and vanishes for symmetry-breaking ones in the low-energy limit.
This behavior is consistent with previous studies~\cite{Kato:2012} and can be explicitly confirmed within the EFT framework.

The symmetry-preserving perturbation corresponds to a spatial modulation of the stiffness $g^s(x)=f_s^2(x)$ and magnetization $\rho(x)$ with $g^t_{ab}=0$ in Eq.~\eqref{eq:type-B-eom}.
In this case, the resulting equation of motion reduces to a one-dimensional scattering problem that is \emph{identical} in form to that for superfluid phonons.
The only difference lies in the dispersion relation (linear for type-A phonons and quadratic for type-B magnons).
Nevertheless, the transmission coefficient takes the same functional form, leading to perfect transmission in the low-energy limit.
In contrast, the symmetry-breaking perturbation introduces the term $\Delta(x)$ in Eq.~\eqref{eq:action-potential-breaking}.
As was shown before, the problem thus reduces to a standard potential scattering problem, leading to perfect reflection in the low-energy limit.

These results are fully consistent with our general criterion.

\sect{Concluding remarks}%
\label{sec:Discussion}%
In this Letter, we established that anomalous tunneling of NG modes is a universal low-energy theorem dictated by symmetry and scaling, providing a unifying principle for tunneling phenomena in systems with spontaneous symmetry breaking.
For symmetry-preserving localized potentials, the transmission coefficient approaches unity in the low-energy limit, whereas symmetry-breaking perturbations lead to perfect reflection.
This universal behavior is determined solely by symmetry and scaling, independent of microscopic details.
We further derive the formula \eqref{eq:low-energy-behavior} for the reflection coefficient at low momentum in the symmetry-preserving case, whose $k^2$ coefficient is determined by the low-energy functions of NG modes.

We illustrate this result with superfluid and ferromagnetic systems, exemplifying type-A and type-B NG modes~\cite{Watanabe:2012,Hidaka:2013}.
Despite their different dispersion relations, both exhibit the same universal low-energy behavior, demonstrating that anomalous tunneling is insensitive to the type of NG mode.
The superfluid case provides an explicit realization of our result, as shown in Fig.~\ref{fig:transmission} and Eq.~\eqref{eq:R-coefficient}.

While we investigated systems with internal symmetry breaking, systems with broken spacetime symmetries or nonequilibrium steady states exhibit richer EFT structures~\cite{Watanabe:2013,Kobayashi:2014,Takahashi:2015,Hidaka:2015,Minami:2018,Hongo:2019,Hidaka:2020,Hongo:2021,Fujii:2022b,Akyuz:2024}.
Exploring whether analogous low-energy theorems govern NG-mode scattering in such systems is an intriguing direction for future research.

\smallskip
\begin{acknowledgments}
The authors thank D.~T.~Son for valuable discussions.
The authors also thank the Tohoku Forum for Creativity at Tohoku University. Discussions during the junior research program ``Universality of Strongly Correlated Few-body and Many-body Quantum Systems" were useful to complete this work.
K.F. is supported by JSPS KAKENHI Grant Numbers JP24KJ0062 and JP26K17064.
D.K. is supported by JSPS KAKENHI Grant Number JP25K17319.
M.H. is supported by the Japan Society for the Promotion of Science (JSPS) KAKENHI Grants No. 23K25870, No. 25K01002, No. 25K07316, and No. 26H01407.
This work was partially supported by the COREnet project of RCNP at The University of Osaka, RIKEN iTHEMS, and Niigata University Quantum Research Center (NU-Q).

\end{acknowledgments}

\bibliography{anomalous-tunneling}

%
%

\end{document}